\documentclass[twocolumn,aps,prd,   
               preprintnumbers,numbers,sort&compress,
               nofootinbib,
               showpacs,
               floatfix,
               colorlinks,
               linkcolor=blue,
               citecolor=blue,
               10pt]{revtex4-1}

\usepackage{graphicx,amsmath,amssymb,bm}
\usepackage{psfrag}
\usepackage{feynmp}
\usepackage{hyperref}
\usepackage{enumitem}

\newcommand{\exclude}[1]{}

\newcommand{\beq}{\begin{equation}}
\newcommand{\eeq}{\end{equation}}
\newcommand{\be}{\begin{eqnarray}}
\newcommand{\ee}{\end{eqnarray}}

\newcommand{\Lqcd}{\Lambda_{\mathrm{QCD}}}

\def\dd{ \,\mathrm{d} }

\def\+{\dagger}
\def\la{\langle}
\def\ra{\rangle}
\def\<{\langle}
\def\>{\rangle}

\begin{document}

\title{Metastable vacuum decay and $\theta$ dependence in gauge theory. Deformed QCD as a toy model.}

\author{Amit Bhoonah, Evan Thomas \& Ariel R. Zhitnitsky} 



\begin{abstract}
We study a number of different ingredients related to the $\theta$ dependence, metastable excited vacuum states and other related subjects using a simplified version of QCD, the so-called ``deformed QCD''.
This model is a weakly coupled gauge theory, which however preserves  all the relevant essential elements allowing us to study hard and nontrivial features which are known to be present in real strongly coupled QCD.
Our main focus in this work is to test the ideas related to the metastable vacuum states (which are known to be present in strongly coupled QCD in large $N$ limit) in a theoretically controllable manner using the ``deformed QCD'' as a toy model.
We explicitly show how the metastable states emerge in  the system, why their life time is large, and why these metastable states must be present in the system for the self-consistency of the entire picture of the QCD vacuum.
We also speculate on possible relevance of the metastable vacuum states in explanation of the violation of local $\cal{P}$ and $\cal{CP}$ symmetries  in heavy ion collisions.  \\
PACS: 11.15-q
\end{abstract}

\maketitle

\section{Introduction and motivation} \label{introduction}

A study of the the QCD vacuum state in the strong coupling regime is the prerogative of numerical Monte Carlo lattice computations.
However, a number of very deep and fundamental questions about the QCD vacuum structure can be addressed and, more importantly, answered using some simplified versions of QCD. 
In the present paper, we study a set of questions related to metastable vacuum states and their decay to the true vacuum state using the so-called ``deformed QCD'' toy model wherein we can work analytically.
This model describes a weakly coupled gauge theory, which however preserves many essential elements expected for true QCD, such as confinement, degenerate topological sectors, proper $\theta$ dependence, etc.
This allows us to study difficult and nontrivial features, particularly related to vacuum structure, in an analytically tractable manner.

The fact that some high energy metastable vacuum states must be present in a gauge theory system in the large $N$ limit has been known for quite some time \cite{Witten:1980sp}.
A similar conclusion  also  follows from the  holographic description of QCD as originally discussed in \cite{wittenflux}. Furthermore, it has been known since  \cite{Shifman:1998if} that the decay rate of these excited vacua in large $N$ limit in strongly coupled pure gauge theory can be estimated as $\Gamma\sim \exp (-N^4)$.

The fundamental observation on the emergence of these excited vacuum states was made in a course of studies related to the resolution of the $U(1)_A$ problem in QCD in the large $N$ limit\cite{witten,ven,vendiv}.
 In the present work we do not introduce quarks (which play an essential role in the formulation of the $U(1)_A$ problem) into the system, but rather, study pure gluodynamics, and the metastable vacuum states which occur there. 
Nevertheless, the key object relevant for the resolution of the $U(1)_A$ problem, the so-called topological susceptibility $\chi$, still emerges in our discussions in pure gluodynamics because it plays an important role in understanding of the spectrum of the ground state and multiple metastable states.
Indeed, the topological susceptibility is defined as $\chi (\theta)=\frac{\partial^2 E_{\rm vac}(\theta)}{\partial \theta^2}$.
Therefore, the information about the ground (or in general metastable) states $E_{\rm vac}(\theta)$ is related to the $\theta$ behaviour of the system formulated in terms of the topological susceptibility $\chi (\theta)$.  

When some deep questions are studied in a simplified version of a theory, there is always a risk that some effects which emerge in the simplified version of the theory could be just artifacts of the approximation, rather than genuine consequences of the original underlying theory.
Our study using the ``deformed QCD'' as a toy model is not free from this potential difficulty with misinterpretation of artifacts as inherent features underlying QCD.
Nevertheless, there are few strong arguments suggesting that we indeed study some intrinsic features of the system rather than some artificial effects.
The first argument is discussed in the original paper on ``deformed QCD'' \cite{Yaffe:2008} where it has been claimed that this model describes a smooth interpolation between strongly coupled QCD and the weakly coupled ``deformed QCD'' without any phase transition.
In addition, there are a few more arguments based on our previous experience with the ``deformed QCD'' model, see below, which also strongly suggest that we indeed study some intrinsic features of QCD rather than some artifact of the deformation.

Our arguments are based on the computation \cite{Thomas:2011ee} of the contact term in the ``deformed QCD'', see also \cite{Unsal:2012zj} with some related discussions.
The key point is that this contact term with a positive sign (in the Euclidean formulation) in the topological susceptibility $\chi$ is required for the resolution of the $U(1)_A$ problem \cite{witten,ven,vendiv}.
At the same time, any physical propagating degrees of freedom must contribute with a negative sign, see \cite{Thomas:2011ee} with details.
In \cite{witten} this positive contact term has been simply postulated while in \cite{ven,vendiv} an unphysical Veneziano ghost was introduced into the system to saturate this term with the ``wrong'' sign in the topological susceptibility.
This entire, very non-trivial picture, has been successfully confirmed by numerical lattice computations.
More importantly for the present studies, this picture has been supported by analytical computations in ``deformed QCD'' in which all the nontrivial crucial elements for the resolution of the $U(1)_A$ problem indeed emerge in analytical analysis. 

Indeed, the non-dispersive contact term in topological susceptibility can be explicitly computed in this model and is given by \cite{Thomas:2011ee}
\be \label{YM}
  \chi_{\rm contact} = \int d^4 x \< q(x) , q(0) \>	\sim  \int d^3 x \left[ \delta(\mathbf{x}) \right], 
\ee
where $q(x)$ is the topological density operator.
It has the required ``wrong sign'' as this contribution is not related to any physical propagating degrees of freedom, but is rather related to the topological structure of the theory, and has a $\delta(\mathbf{x})$ function structure as it should.
In this model $\chi$  is saturated by fractionally charged weakly interacting monopoles describing the tunnelling transitions between topologically distinct, but physically equivalent topological winding sectors.
Furthermore, the $\delta (\mathbf{x})$ function in (\ref{YM}) should be understood as total divergence related to the infrared (IR) physics, rather than to ultraviolet (UV) behaviour as explained in \cite{Thomas:2011ee}
\be	\label{divergence}
  \chi_{\rm contact} \sim   \int \delta (\mathbf{x})  \dd^3x  = \int   \dd^3x~
  \partial_{\mu}\left(\frac{x^{\mu}}{4\pi x^3}\right). 
\ee
The singular behaviour of the contact term has been confirmed by the lattice computations where it has been found that the singular behaviour at $x \rightarrow 0$ is an inherent IR feature of the underlying QCD rather than some lattice size effect \cite{Horvath:2005cv,Ilgenfritz:2007xu,Ilgenfritz:2008ia,Bruckmann:2011ve}.

In addition, one can explicitly see how the Veneziano ghost postulated in \cite{ven,vendiv} is explicitly expressed in terms of auxiliary topological fields which saturate the contact term ({\ref{YM}) in this model \cite{Zhitnitsky:2013hs}.
In other words, the $\eta'$ field in this model generates its mass (which is precisely the formulation of the $U(1)_A$ problem) as a result of a mixture of the Goldstone field with the topological auxiliary field governed by a Chern-Simons like action, see \cite{Zhitnitsky:2013hs} for the details.

All these features related to the $\theta$ dependence which are known to be present in strongly coupled regime also emerge in the weakly coupled ``deformed QCD'' toy model.
Therefore, we interpret such behaviour as a strong argument supporting our assumption that the ``deformed QCD'' model properly describes, at least qualitatively, the features related to the $\theta$ dependence and vacuum structure of QCD, including the presence of metastable states which is main subject of the present work.

The specific computations we perform related to the metastable vacuum states have never been performed using numerical lattice (or any other) methods. 
Therefore, we do not have the same level of luxury present in our previous studies of the contact term \cite{Thomas:2011ee} in which our results were supported by numerous lattice computations.
Nevertheless, as the specific questions about the metastable states are closely related to much more generic studies of the $\theta$ dependence in the system, as reviewed above, we are still confident that our results presented below, based on the ``deformed QCD'' model, are inherent qualitative properties of QCD rather than some artificial effects which may occur due to the deformation. 

Our presentation is organized as follows.
We start in section \ref{deformedqcd} by reviewing a simplified (``deformed'') version of QCD which, on one hand, is a weakly coupled gauge theory wherein computations can be performed in theoretically controllable manner.
On other hand, this deformation preserves all the elements relevant to our study such as confinement, degeneracy of topological sectors, nontrivial $\theta$ dependence, presence of non-dispersive contribution to topological susceptibility, and other crucial aspects pertinent to the study of the metastable states.
In section \ref{classification} we explicitly demonstrate the presence of metastable states in this model. In section \ref{Coleman} we review the general strategy to compute a decay of metastable vacuum states to the true vacuum in the path integral formulation.
Finally, in section \ref{computations} we present our numerical analysis on the life time of the metastable states as a function of a ``semi-classicality'' which is a parameter determining the region of validity of our semiclassical computations.
We conclude in section \ref{conclusion} with speculations on possible consequences and manifestations of our results for physics of heavy ion collisions where a metastable state might be formed as a result of collision, and the system, which is order the size of a nuclei, might be locked in this state for sufficiently long period of time $\sim$ 10 fm/c.

\section{Deformed QCD} \label{deformedqcd}

Here we overview the ``center-stabilized'' deformed Yang-Mills developed in \cite {Yaffe:2008}.
In the deformed theory an extra ``deformation'' term is put into the Lagrangian in order to prevent the center symmetry breaking that characterizes the QCD phase transition between ``confined'' hadronic matter and ``deconfined'' quark-gluon plasma, thereby explicitly preventing that transition.
Basically the extra term describes a potential for the order parameter 
The basics of this model are reviewed in section \ref{model}, while in section \ref{classification} we classify the metastable states which is inherent element of the system.

\subsection{The Model}\label{model}

We start with pure Yang-Mills (gluodynamics) with gauge group $SU(N)$ on the manifold $\mathbb{R}^{3} \times S^{1}$ with the standard action
\be \label{standardYM}
	S^{YM} = \int_{\mathbb{R}^{3} \times S^{1}} d^{4}x\; \frac{1}{2 g^2} \mathrm{tr} \left[ F_{\mu\nu}^{2} (x) \right],
\ee
and add to it a deformation action,
\be \label{deformation}
	\Delta S \equiv \int_{\mathbb{R}^{3}}d^{3}x \; \frac{1}{L^{3}} P \left[ \Omega(\mathbf{x}) \right],
\ee 
built out of the Wilson loop (Polyakov loop) wrapping the compact dimension
\be \label{loop}
	\Omega(\mathbf{x}) \equiv \mathcal{P} \left[ e^{i \oint dx_{4} \; A_{4} (\mathbf{x},x_{4})} \right].
\ee
The parameter  $L$ here is the length of the compactified dimension which is assumed to be small. 
The coefficients of the polynomial $P \left[ \Omega(\mathbf{x}) \right]$ can be suitably chosen such that the deformation potential (\ref{deformation}) forces unbroken symmetry at any compactification scales.
At small compactification $L$ the gauge coupling is small so that the semiclassical computations are under complete theoretical control \cite{Yaffe:2008}.

As described in \cite{Yaffe:2008}, the proper infrared description of the theory is a dilute gas of $N$ types of monopoles, characterized by their magnetic charges, which are proportional to the simple roots and affine root $\alpha_{a} \in \Delta_{\mathrm{aff}}$ of the Lie algebra for the gauge group $U(1)^{N}$.
For a fundamental monopole with magnetic charge $\alpha_{a} \in \Delta_{\mathrm{aff}}$ (the affine root system), the topological charge is given by
\be \label{topologicalcharge}
	Q = \int_{\mathbb{R}^{3} \times S^{1}} d^{4}x \; \frac{1}{16 \pi^{2}} \mathrm{tr} \left[ F_{\mu\nu} \tilde{F}^{\mu\nu} \right]
		= \pm\frac{1}{N},
\ee
and the Yang-Mills action is given by
\be \label{YMaction}
	S_{YM} = \int_{\mathbb{R}^{3} \times S^{1}} d^{4}x \; \frac{1}{2 g^{2}} \mathrm{tr} \left[ F_{\mu\nu}^{2} \right] = \frac{8 \pi^{2}}{g^{2}} \left| Q \right|.
\ee
The $\theta$-parameter in the Yang-Mills action can be included in conventional way,
\be \label{thetaincluded}
	S_{\mathrm{YM}} \rightarrow S_{\mathrm{YM}} + i \theta \int_{\mathbb{R}^{3} \times S^{1}} d^{4}x\frac{1}{16 \pi^{2}} \mathrm{tr}
		\left[ F_{\mu\nu} \tilde{F}^{\mu\nu} \right],
\ee
with $\tilde{F}^{\mu\nu} \equiv \epsilon^{\mu\nu\rho\sigma} F_{\rho\sigma}$.

The system of interacting monopoles, including the $\theta$ parameter, can be represented in the dual sine-Gordon form as follows \cite{Yaffe:2008,Thomas:2011ee},
\be
\label{thetaaction}
	S_{\mathrm{dual}} &=& \int_{\mathbb{R}^{3}}  d^{3}x \frac{1}{2 L} \left( \frac{g}{2 \pi} \right)^{2}
		\left( \nabla \bm{\sigma} \right)^{2} \nonumber \\
	    &-& \zeta  \int_{\mathbb{R}^{3}}  d^{3}x \sum_{a = 1}^{N} \cos \left( \alpha_{a} \cdot \bm{\sigma}
		+ \frac{\theta}{N} \right), 
\ee
where $\zeta$ is magnetic monopole fugacity which can be explicitly computed in this model using the conventional semiclassical approximation.
The $\theta$ parameter enters the effective Lagrangian (\ref{thetaaction}) as $\theta/N$ which is the direct consequence of the fractional topological charges of the monopoles (\ref{topologicalcharge}).
Nevertheless, the theory is still $2\pi$ periodic.
This $2\pi$ periodicity of the theory is restored not due to the $2\pi$ periodicity of Lagrangian (\ref{thetaaction}).
Rather, it is restored as a result of summation over all branches of the theory when the levels cross at $\theta=\pi (mod ~2\pi)$ and one branch replaces another and becomes the lowest energy state as discussed in \cite{Thomas:2011ee}.
 
Finally, the dimensional parameter which governs the dynamics of the problem is the Debye correlation length of the monopole's gas, 
\be \label{sigmamass}
	m_{\sigma}^{2} \equiv L \zeta \left( \frac{4\pi}{g} \right)^{2}.
\ee
The average number of monopoles in a ``Debye volume'' is given by
\begin{equation} \label{debye}
	{\cal{N}}\equiv	m_{\sigma}^{-3} \zeta = \left( \frac{g}{4\pi} \right)^{3} \frac{1}{\sqrt{L^3 \zeta}} \gg 1,
\end{equation} 
The last inequality holds since the monopole fugacity is exponentially suppressed, $\zeta \sim e^{-1/g^2}$, and in fact we can view (\ref{debye}) as a constraint on the region validity where semiclassical approximation is justified. This parameter ${\cal N}$ is therefore one measure of ``semi-classicality''.

For our studies in what follows it is convenient to express  the action  in terms of dimensionless variables.
We rescale $x$ as follows  $x= x'/m_{\sigma}$ such that  $x'$ becomes a  dimensionless coordinate.  All distances now are measured in units of $m^{-1}_{\sigma}$. With this rescaling the potential term is explicitly proportional to the parameter of semi-classicality ${\cal N} $. 
 The coefficient on the kinetic term in the above action (\ref{thetaaction}) also esquires the same factor ${\cal N} $ such that the action   (\ref{thetaaction}) assumes  a very nice form:
\be	\label{action}
	S &=& {\cal N} \int_{\mathbb{R}^{3}} d^{3}x \sum_{n=1}^{N}\frac{1}{2}\left( \nabla \sigma_n \right)^{2}
	      \nonumber \\
	  &-& {\cal N} \int_{\mathbb{R}^{3}}  d^{3}x \sum_{a = 1}^{N} \cos \left(\sigma_{n} - \sigma_{n+1} 
	      + \frac{\theta}{N} \right), 
\ee 
with $\sigma_{N+1}$ identified with $\sigma_{1}$.  In formula (\ref{action}) we used $x$ rather than $x'$ to simplify notations.  
The Lagrangian entering the action (\ref{action}) is then dimensionless  with a prefactor ${\cal N}$, such that the rest of the action depends on $N$ (in the number of fields), but no longer depends on $g$, the gauge coupling, or $L$, the compactification scale.
This is the form of the action we use in our calculations.
Note that   the large $N$ limit in the so called ``double scaling'' limit had been discussed previously in \cite{Yaffe:2008}. It has been argued that 
in this limit we can consider any finite $N$ but cannot consider the formal limit $N \rightarrow \infty$ since the parameter space shrinks to  a point. This will not matter for the present work as we always carry out  the computations for large, but finite $N$.  

\subsection{Metastable Vacuum States}\label{classification}

Here we concentrate on the Euclidean potential density for the $\sigma$ fields at $\theta=0$,
\be	\label{potential}
	U(\bm{\sigma}) = \sum_{n=1}^{N} \left[ 1 - \cos \left( \sigma_{n} - \sigma_{n+1} \right) \right],
\ee
where again $\sigma_{N+1}$ is identified with $\sigma_{1}$. To simplify notations we skip a large common 
factor $\cal{N}$ in our discussions which follow. We restore this factor in our final formula.  
Also, we have added a constant ($N$) so that the potential is positive semi-definite. The lowest energy state, denoted by $\bm{\sigma}^{(-)}$, is the state with all $\sigma$ fields sitting at the same value ($\sigma_n = \sigma_{n+1}$) and has zero energy.
This is clearly the true ground state of the system, but there are also potentially some higher energy metastable states.
For an extremal state we must have
\be
	\frac{\partial U}{\partial \sigma_n} = 0
\ee
for all $n$, which gives immediately
\be	\label{necessary}
	\sin \left( \sigma_{n} - \sigma_{n+1} \right) = \sin \left( \sigma_{n-1} - \sigma_{n} \right).
\ee
A necessary condition for a higher energy minimum of the potential is thus that the $\sigma$ fields are evenly spaced around the unit circle or (up to a total rotation),
\be	\label{sigmanecessary}
	\sigma_n = m \frac{2 \pi n}{N},
\ee
where $m$ is an integer.
A sufficient condition is then
\be
\label{derivative}
	\frac{\partial^2 U}{\partial \sigma_n^2} > 0,
\ee
again for all $n$. This gives us
\be
	\cos \left( \sigma_{n} - \sigma_{n+1} \right) + \cos \left( \sigma_{n-1} - \sigma_{n} \right) > 0,
\ee
which using (\ref{sigmanecessary}) gives
\be	\label{sufficient}
	\cos \left( m \frac{2 \pi}{N} \right) > 0.
\ee
So, we get a constraint on $m$ in the form of (\ref{sufficient}), and also on $N$.
From (\ref{sufficient}) it is quite obvious that metastable states always  exist for sufficiently large $N$, which is is definitely consistent with old and very generic arguments \cite{Witten:1980sp}. In our simplified version of the theory one can explicitly see how these metastable states emerge in the system, and how they are classified
in terms of the scalar magnetic potential fields  $\bm{\sigma} (\mathbf{x})$.

One should also remark here that a non-trivial solution with $m\neq 0$ in (\ref{sufficient}) does not exist\footnote{$N=4$ deserves a special consideration as at $m=\pm1$ the second derivative (\ref{derivative}) vanishes.
It may imply a presence of the massless particles in the spectrum for these excited vacuum states.
It may also correspond to a saddle point in configuration space.
We shall not elaborate on this matter in present work.} in this simplified model for the lowest $N = 2, 3, 4$.
Therefore, in our study we always assume $N \geq 5$. 

Looking back at the potential (\ref{potential}), the lowest energy of the possibilities are given by $m = \pm1$, so that the lowest energy metastable states, denoted by $\bm{\sigma}^{(+)}$, are given by (again up to a constant rotation)
\be	\label{metasigma}
	\sigma_n^{(+)} = \pm \frac{2 \pi n}{N}.
\ee

To understand the physical meaning of the solutions describing the nontrivial metastable vacuum states, we recall that the operator $e^{i \alpha_{a} \cdot \bm{\sigma} (\mathbf{x})}$ is the creation operator for a monopole of type $a$ at point $\mathbf{x}$, as it was explicitly demonstrated in \cite{Thomas:2011ee},
\be
\label{operator}
{\cal{M}}_a (\mathbf{x}) =e^{i \alpha_{a} \cdot \bm{\sigma} (\mathbf{x})}.
\ee
Therefore, the vacuum expectation value $\langle{\cal{M}}_a (\mathbf{x})\rangle$ describes the magnetization of a given metastable ground state classified by the parameter $m$.
As one can see from (\ref{sigmanecessary}), the corresponding vacuum expectation value $\langle{\cal{M}}_a (\mathbf{x})\rangle$ always assumes the element from the centre of the $SU(N)$ group.
Specifically, for the lowest metastable vacuum states given by (\ref{metasigma}),
the magnetization is given by
\be
\label{magnetization}
\langle{\cal{M}}_a (\mathbf{x})\rangle =\exp{\left[{\pm i\frac{2\pi}{N}}\right]}.
\ee 
The fact that the confinement in this model is due to the condensation of fractionally charged monopoles has been known since the original paper \cite{Yaffe:2008}.
Now we understand the structure of the excited metastable states also; mainly, these metastable vacuum states can be also thought of as a condensate of the monopoles.
However, the condensates of different monopole types, $n$ from (\ref{metasigma}), are now shifted by a phase such that the corresponding  magnetization receives a non-trivial phase (\ref{magnetization}).    

A different, but equivalent way to classify all these new metastable vacuum states is to compute the expectation values for the topological density operator for those states. By definition
\be	\label{top}
  \langle \frac{1}{16 \pi^{2}} \mathrm{tr} \left[ F_{\mu\nu} \tilde{F}^{\mu\nu} \right]\rangle_m
    \equiv-i\frac{\partial S_{\rm dual}(\theta)}{\partial \theta}|_{\theta=0} \\
  =-i\frac{\zeta}{L}\langle\sin \left(\alpha_{a} \cdot \bm{\sigma}\right)\rangle_m=-i\frac{\zeta}{L}\sin \left(\frac{2\pi m}{N}\right).\nonumber
\ee
The imaginary $i$ in this expression should not confuse the readers as we work in the Euclidean space-time.
In Minkowski space-time this expectation value is obviously a real number.
A similar phenomenon is known to occur in the exactly solvable two dimensional Schwinger model wherein the expectation value for the electric field in the Euclidean space-time has an $i$, see \cite{Zhitnitsky:2013hba} for discussions within  present context. 
The expectation value (\ref{top}) is the order parameter of a given metastable state. 

The crucial point we want to make here is that a metastable vacuum state with $m \neq 0$ in general violates $\cal{P}$ and $\cal{CP}$ invariance since the topological density operator itself is not invariant under these symmetries.
Precisely this observation inspires our suggestion, to be discussed in conclusion, that such metastable states could be the major source of the local $\cal{P}$ and $\cal{CP}$ violation observed in heavy ion collisions at the Relativistic Heavy Ion Collider (RHIC) at Brookhaven, and the Large Hadron Collider (LHC). 

Now we come back to our discussions of the lowest metastable states (\ref{metasigma}).
Putting the metastable configuration back into the potential (\ref{potential}) we find that the energy density separation between the true ground state and the lowest metastable states (\ref{metasigma}) is given by\footnote{\label{N}One should comment here that the vacuum energy of the ground state $E^{(\pm)}\sim N$ in this model scales as $N$ in contrast with conventional $N^2$ scaling in strongly coupled QCD.
However, the ratio ${\epsilon}/{E^{(\pm)}} \sim  N^{-2}$ shows the same scaling as in strongly coupled QCD.
The difference in behaviour in large $N$ limit between weakly coupled ``deformed QCD" and strongly coupled QCD obviously implies that  one should anticipate a different asymptotic scaling  for the decay rate in large $N$ limit  in our simplified model in comparison with result \cite{Shifman:1998if}. As we discuss in sections \ref{results},\ref{improved} this is indeed the case. 

Furthermore, the region of validity in this model shrinks to a point in the limit $N\rightarrow\infty$ as discussed in \cite{Yaffe:2008}. Therefore, the asymptotical behaviour at $N\rightarrow\infty$ should be considered  with great caution.}
\be	\label{epsilon}
	\epsilon \equiv \left (E^{(+)}-E^{(-)}\right) = N \left[ 1 - \cos \left( \frac{2 \pi}{N} \right) \right]. 
\ee
The choice of sign in (\ref{metasigma}) is irrelevant for the purposes of calculating the vacuum decay since the two states $m = \pm 1$ are degenerate in terms of energy and have the same energy splitting with respect to the ground state.
These states, however, are physically distinct as the expectation value of the gauge invariant operator (\ref{top}) has opposite signs for these two metastable vacuum states.
This implies that all $\cal{P}$ and $\cal{CP}$ effects will have the opposite signs for these two states, while the probability to form these two metastable states is identical, as is the decay rate.
So, while our fundamental Lagrangian is invariant under these symmetries, the metastable vacuum states, if formed, may spontaneously break that symmetry.
 
\section{Metastable Vacuum Decay}\label{Coleman}

In this section we briefly review the general theory and framework for calculating metastable vacuum decay rates in Quantum Field Theory, restating the important results for the three dimensional model discussed above.
For a more thorough discussion see \cite{Kobzarev:1974cp, Coleman:1977sc, Coleman:1977qc}.
The process for the decay of a metastable vacuum state to the true vacuum state is analogous to a bubble nucleation process in statistical physics.
Considering a fluid phase around the vaporization point, thermal fluctuations will cause bubbles of vapor to form.
If the system is heated beyond the vaporization point, the vapor phase becomes the true ground state for the system.
Then, the energy gained by the bulk of a bubble transitioning to the the vapor phase goes like a volume while the energy cost for forming a surface (basically a domain wall) goes like an area.
Thus, there is some critical size such that smaller bubbles represent a net cost in energy and will collapse while larger bubbles represent a net gain in energy.
Once a bubble forms which is larger than the critical size it will grow to consume the entire volume and transition the whole of the sample to the vapor phase.
To understand the lifetime of such a 'superheated' liquid state, the important calculation is, therefore, the rate of nucleation of critical bubbles per unit time per unit volume ($\Gamma/V$).
Similarly, we aim to calculate this decay rate for our system with from the metastable state $\bm{\sigma}^{(+)}$ to the ground state $\bm{\sigma}^{(-)}$, though through quantum rather than thermal fluctuations.
Classically, a system in the configuration $\bm{\sigma}^{(+)}$ is stable, but quantum mechanically the system is rendered unstable through barrier penetration (tunneling).

The semiclassical expression for the tunneling rate per unit volume is given by \cite{Kobzarev:1974cp, Coleman:1977sc, Coleman:1977qc}
\be	\label{semiclassical}
	\frac{\Gamma}{V} = A e^{-S_E(\bm{\sigma}_b)/\hbar} \left[ 1 + O \left( \hbar \right) \right]
\ee
where $S_E$ is the Euclidean action (\ref{action}) and is evaluated in the field configuration called the ``Euclidean bounce'' which we have denoted $\bm{\sigma}_b$.
The Euclidean bounce is a finite action, spherically symmetric configuration which solves the classical equations of motion and interpolates from the metastable state to a configuration ``near'' the ground state and back.

In the limit of small separation energy $\epsilon$ the bounce approaches $\bm{\sigma}^{(-)}$ more closely and spends longer in the region nearby, so that the bounce configuration resembles a bubble with the interior at $\bm{\sigma}^{(-)}$, the exterior at $\bm{\sigma}^{(+)}$, and a domain wall surface interpolating between the two\footnote{One should comment here that this model also exhibits very different types of the domain walls considered in \cite{Thomas:2012tu}. The objects discussed in  \cite{Thomas:2012tu} are fundamentally different from solutions discussed in the present work as they essentially describe the tunnelling events between different topological sectors, while in the present work the domain wall-like objects play the auxiliary role in order to evaluate the life time of a metastable state.}.
If the bubble is very large, corresponding to very small $\epsilon$, then the curvature at the interpolating surface is small and the surface appears flat.

Therefore, if the separation energy, $\epsilon$, between the two states is small, we need only solve for the one dimensional soliton interpolating between $\bm{\sigma}^{(+)}$ and $\bm{\sigma}^{(-)}$ which solves
\be	\label{1daction}
	S_1 \! = \!\! \int \!\! dx \! \sum_{n=1}^{N}\left[\frac{1}{2}\left( \frac{d\sigma_n}{dx} \right)^{2} \!\! + 1 - \cos \left(\sigma_{n} - \sigma_{n+1} \right) \right]\!\!.
\ee
This is called the thin-wall approximation, and is the framework in which we will work.
In the deformed model, as discussed in the previous section, the separation $\epsilon \sim 1/N$, so that the thin-wall approximation coincides with the large $N$ approximation.

For the thin wall approximation the full action action reduces to
\be	\label{3daction}
	S_3 = 4 \pi R^2 S_1 - \frac{4}{3} \pi R^3 \epsilon = \frac{16}{3} \pi \frac{S_1^3}{\epsilon^2},
\ee
where the last step is computed by using variational analysis to get $R = 2 S_1/\epsilon$.
Notice again the similarity to a bubble nucleation problem.
This extremal action with respect to the bubble size is in fact a maximum, and as such the action increases with $R$ for smaller size and decreases with $R$ for larger.
Hence, the bounce configuration which saturates the decay rate is essentially a bubble of critical size as discussed when making this analogy to bubble nucleation.  

The condition for the validity of the thin wall approximation is essentially that the interior of the bubble is very near the true ground state $\bm{\sigma}^{(-)}$ so that it is nearly stable and stays near $\bm{\sigma}^{(-)}$ for large $\rho$.
We want $R\mu \gg 1$, where $\mu^2 = \partial^2U/\partial\sigma_n^2(\bm{\sigma}^{(-)})$ is the curvature of the potential at the ground state; here $\mu = \sqrt{2}$.
Thus, we need
\be	\label{condition}
	2\sqrt{2}S_1 \gg \epsilon,
\ee
where $\epsilon$ is given by (\ref{epsilon}).

We now have everything required to calculate the exponent for the vacuum decay (\ref{semiclassical}) assuming we can solve for a classical path associated with the one dimensional action (\ref{1daction}) interpolating between the two states $\bm{\sigma}^{(+)}$ and $\bm{\sigma}^{(-)}$.
We have not discussed the coefficient $A$, and indeed it is a much more complicated problem related to the functional determinant of the full differential operator, $\delta^2S/\delta\sigma^2$.
This calculation is beyond the scope of this work for the deformed model as it does not change 
the basic physical picture advocating in this work. We want to see that the leading  factor  in the decay rate is indeed exponentially small, and our computations are justified  as long as our semiclassical parameter (\ref{debye}) is sufficiently large, ${\cal{N}}\gg 1$.  
Furthermore, we anticipate that the dependence on $N$  in the exponent is much more important than the  $N$ dependence in  pre exponential factor $\sim \delta^2S/\delta\sigma^2$. 
The only power -like corrections which may emerge    from the determinant is through a factor of $\sqrt{S_E/2\pi}$ for each zero mode \cite{Coleman:1977sc}, and we can safely neglect these corrections  in comparison with much more profound exponential behaviour in $N$, see below sections \ref{results}, \ref{improved}.

\section{Computations}\label{computations}

We now proceed to solve the equations of motion
\be	\label{EOM}
	\frac{d^2\sigma_n}{dx^2} = \sin\left(\sigma_n-\sigma_{n+1}\right) - \sin\left(\sigma_{n-1}-\sigma_n\right),
\ee
with $\sigma_{N+1}$ identified with $\sigma_1$, derived from the action (\ref{1daction}) subject to the boundary conditions
\be	\label{leftboundary}
	\sigma_n \left( x \rightarrow -\infty \right) = 0,
\ee
and
\be	\label{rightboundary}
	\sigma_n \left( x \rightarrow +\infty \right) = \frac{2 \pi}{N}n + \varphi.
\ee
The $\varphi$ in (\ref{rightboundary}) is a relative rotation angle between the two boundaries, since each of the two states are only defined up to a rotation. 
The angle is determined by demanding a minimal action interpolation.
That is, we should minimize the action with respect to the interpolating field configuration and also with respect to this angle $\varphi$.
The final solution thus obtained will then be defined only up to an arbitrary total rotation which will be important later.
Additionally, we expect the solution to be a soliton (instanton-like) in the sense that it should be well contained with only exponential tails away from the center so that the interpolation occurs in an exponentially small region.
The characteristic size of this region, we expect, is given by $m_{\sigma}^{-1}$ in the original model, or just $1$ in dimensionless notations used here.
Then, we can calculate the vacuum decay rate as
\be	\label{decay}
	\frac{\Gamma}{V} \sim \left(\frac{S_3}{2\pi}\right)^3 e^{-S_3},
\ee
where we have put in the part of the coefficient that we can calculate related to the zero modes in the system.
There are six zero modes: three translations, two spacial rotations, and the one global $\sigma$-rotation discussed earlier.
We now discuss in section \ref{numerical} the numerical technique employed to solve this problem, and in section \ref{results} the results of these numerical calculations.

\subsection{Numerical Technique}	\label{numerical}


Numerical techniques can be broadly classified as Iterative or Explicit Finite Difference (EFD).
EFD methods involve approximating the derivatives as an N $\times$ N matrix and then inverting the latter only once to find the solution.
They work well for simple linear problems where there is a unique non-trivial solution to the equations of motion.
In iterative methods, we first define an initial guess for the solution and then relax that solution until the residual is below a certain user defined precision.
This is a better technique for non-linear systems where several non-trivial solutions may exist, and is therefore the method we employ here.
The equations of motion (\ref{EOM}) are a coupled version of the sine-Gordon equation and are quite non-linear.

The sine-Gordon equation for a single field, $u^{\prime\prime}=\sin(u)$, has a soliton solution given by 
\be	\label{sgsoliton}
	u\left(x\right) = 4\arctan\left(e^x\right),
\ee
interpolating between $0$ and $2\pi$, which seems like good starting point. 
As such, we choose a similar form for the initial guess at the solution for the coupled equations, and hence define our initial guess to be of the form
\be	\label{guess}
	\sigma_n = \left(\frac{n}{N}+\frac{\varphi}{2\pi}\right)4\arctan\left(e^x\right).
\ee
This initial guess has two important properties; it satisfies the boundary conditions (\ref{leftboundary}) and (\ref{rightboundary}), and tails off toward those boundaries as decaying exponentials for $x \rightarrow \pm \infty$, which is the type of behaviour expected, as discussed previously.

The equations of motion (\ref{EOM}) are on an infinite domain and must be truncated to be solved numerically.
We want to truncate the domain to a region beyond which changes in the $\sigma_i\left(r\right)$ are numerically insignificant.
Given that the tails of the $\sigma_n\left(r\right)$ (and we expect the final solution) are decaying exponentials, choosing the domain $\left[-16,16\right]$ means that the boundary values are within $\sim 10^{-7}$ of their final values and is suitable for our purpose. 

In order to promote numerical stability particularly around the boundary values we employ Chebyshev spectral methods for integrals and derivatives, as described in \cite{Boyd:2001}\cite{Trefethen:2000}, using an unevenly spaced Chebyshev grid given by
\be	\label{chebyshev}
	x_i = cos\left(\frac{\pi i}{N_p}\right), \enspace \forall \enspace 0 \leq i \leq N_p
\ee
where $N_p$ is the number of grid points.
Notice that the Chebyshev grid defines a domain $\left[-1,1\right]$ and so we scale $x \rightarrow \enspace \frac{x}{16}$ in order to express functions with the chosen domain on the spectral grid.

From now on, we will use the following notation: $\sigma_n^i$ denotes the $i^{th}$ grid point of the $n^{th}$ field, where $N$ is the total number of fields while $N_p$ is the number of grid points.
The differentiation matrix ($N_p \times N_p$) is given by \cite{Boyd:2001} (page 570) as
\begin{equation}	\label{matrix}
    D_{ij} = 
    \begin{cases}
	\begin{array}{cc}
		\frac{2N_p^2 + 1}{6} & i=j=0 \\
		-\frac{2N_p^2 + 1}{6} & i=j=N_p \\
		-\frac{x_j}{2\left(1-x_j^2\right)} & 0<i=j<N_p \\
		\frac{\left(-1\right)^{i+j}p_i}{p_j\left(x_i-x_j\right)} & i \ne j
	\end{array}
    \end{cases}
\end{equation}

where,

\begin{equation*}
    p_j = 
    \begin{cases}
	\begin{array}{cc}
	    2 & j = 0 \: or \: N \\
	    1 & \textit{otherwise.} 
	\end{array}
    \end{cases}
\end{equation*}

Any higher derivative is then given by repeated multiplication by $D$.
This differentiation matrix (\ref{matrix}) is basically just the linear operator describing interpolating a function on the grid points by an $N_p^{th}$ order polynomial and differentiating that polynomial.
Since it uses knowledge of the entire function rather than just the few nearby points like a finite difference, the accuracy of the derivative is generally much better than any small order finite difference.
Furthermore, using a grid spaced in this way provides much more numerical stability, counteracting the Runge phenomenon.\cite{Trefethen:2000}


The algorithm we employ to minimize the action with respect to the field configuration is essentially to treat the action as a potential over the configuration space formed by each $\sigma$ field at each grid point, then to take steps in the negative gradient direction.
Essentially, iterating the expression
\be	\label{step}
	\sigma_n^i \rightarrow \sigma_n^i &-& \delta\frac{dS_1}{d\sigma_n^i} \\
	\sigma_n^i \rightarrow \sigma_n^i &+& \delta\left(D^2\sigma_n\right)^i-\delta\sin\left(\sigma_n^i-\sigma_{n+1}^i\right)\nonumber\\
					   &+& \delta\sin\left(\sigma_{n-1}^i-\sigma_n^i\right) 
\ee
where $\delta$ is a chosen step size, which we start as $\delta = 1$.
At each iteration, we enforce the boundary conditions and check if the action (\ref{1daction}) applied to the new configuration is in fact smaller than the old configuration.
If so, we move to the new configuration.
If not, the step was too large and we have overstepped the section of the potential with a downward slope, so we go back to the old configuration and reduce the step size by $\delta \rightarrow \delta/2$ and iterate this procedure until we find a good step or reduce the step size below our desired precision.
We then reset $\delta = 1$ and continue until we cannot find a good step within our desired precision.
Once reaching a position from which no step reduces the action, we are within the defined precision of a minimum of the action.
In order to probe more of the configuration space we took a Monte-Carlo-like approach wherein we adjusted our initial guess by adding some Gaussian noise in an envelope, $(1-\left|x\right|)^2$.
We chose an envelope of this form because we expect that the solution has the sort of exponential tails of our initial guess (\ref{guess}), while we do not know the form of the core of the domain wall.
Thus, it is sensible to probe more of the configuration space related to the specific details of the core.

\subsection{Results}		\label{results}

The first issue we address is the question about the favoured angle, $\varphi$, between the two boundaries, (\ref{leftboundary}) and (\ref{rightboundary}).
In order to find the angle we chose (arbitrarily) $N = 7$ and varied the angle in the range $\left[-\pi,\pi\right]-8\pi/7$, and look at the action $S_1$ as a function of $\varphi$.
The results of that simulation are plotted in figure \ref{phiplot}.
The center point on the plot, $-8\pi/7$, may seem odd, but it is the value for $\varphi$ which leads to a maximally symmetric solution, so it is very believable minimum for the potential, and indeed this is what we see.
\begin{figure}[t]
  \begin{center} 
    \includegraphics[width = .5\textwidth]{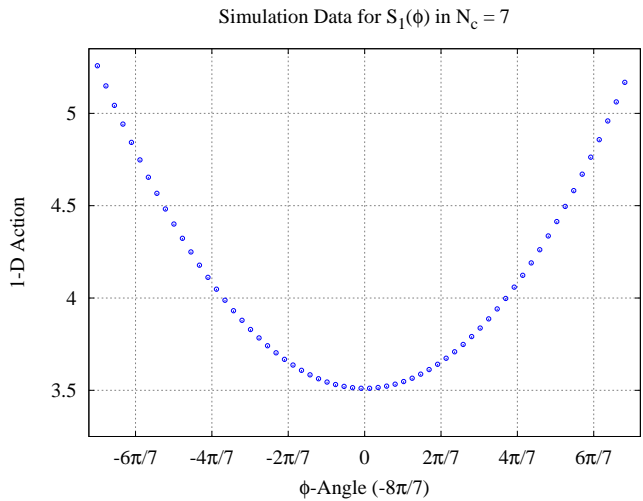}
      \caption{		\label{phiplot}
	A plot of some simulation data for the one dimensional action (\ref{1daction}) as a function of the angle $\varphi$ between the boundary conditions done for $N = 7$.
      }
  \end{center}
\end{figure}
The solution for the minimal $\sigma$ field configuration corresponding to $\varphi = -8\pi/7$ is shown in figure \ref{sigmaplot} across the domain wall.
\begin{figure}[t]
  \begin{center} 
    \includegraphics[width = 0.5\textwidth]{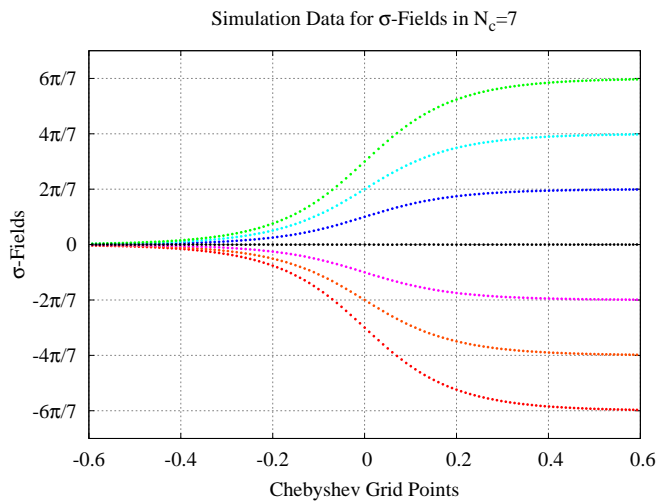}
      \caption{		\label{sigmaplot}
	A plot of some simulation data for the $\sigma$ field configuration plotted across the domain wall done for $N = 7$.
      }
  \end{center}
\end{figure}
\begin{figure*}
  \begin{center} 
    \includegraphics[width = \textwidth]{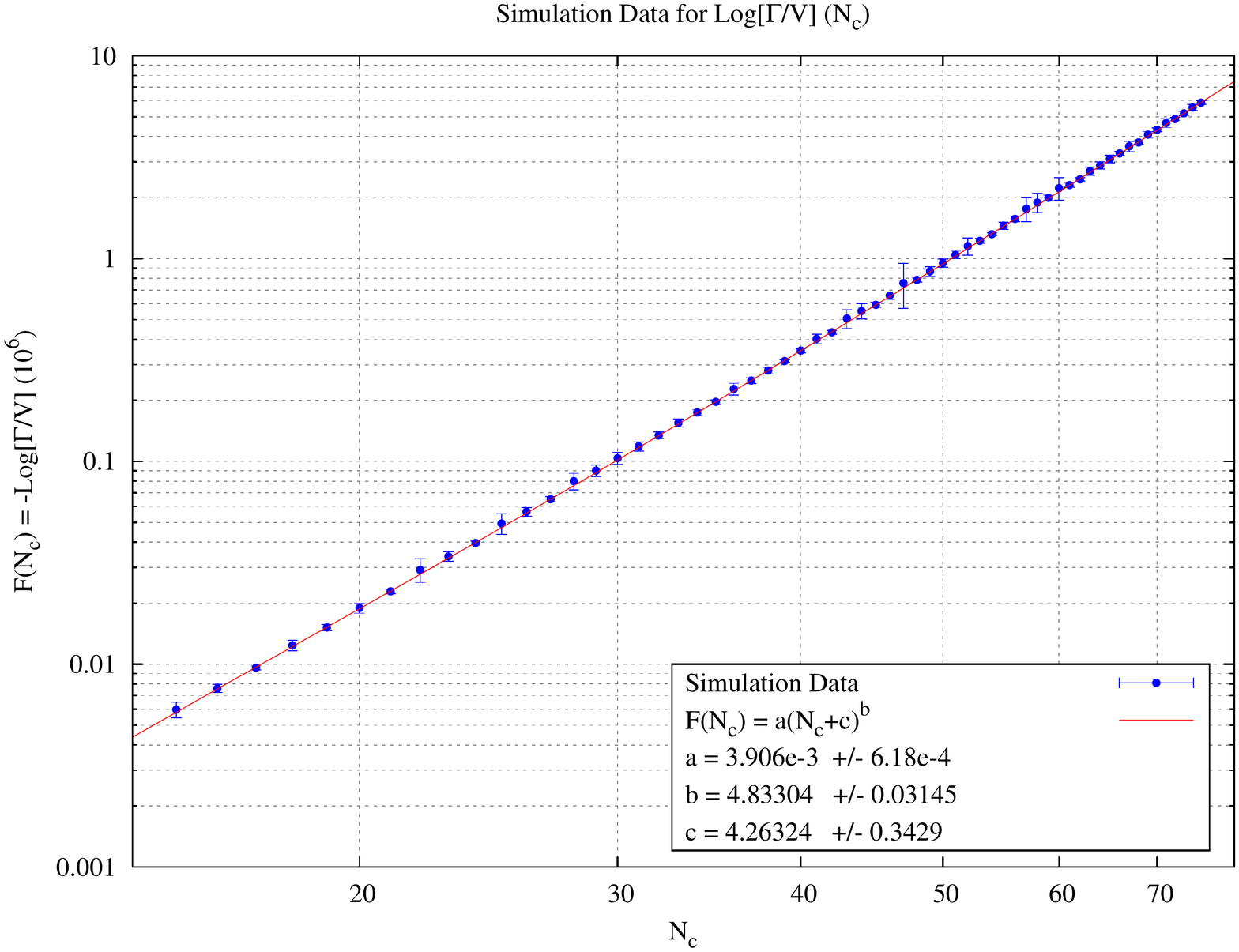}
      \caption{		\label{Fplot}
	A plot of some simulation data for the decay exponent $F\left(N\right)$ plotted for $N$ in the range $15$ to $75$.
      }
  \end{center}
\end{figure*}
Extending these results to arbitrary $N$ we set
\be	\label{phi}
  \varphi = -\pi\left(\frac{N+1}{N}\right),
\ee
which just ensures that the solution we look for is maximally symmetric in the same sense as the fields in figure \ref{sigmaplot}, basically that $\sigma_n = -\sigma_{N+1-n}$.
We have checked that this choice of $\varphi$, (\ref{phi}), does in fact lead to the lowest action configuration for $N = 20$ and $N = 35$ with results much like those shown in figure \ref{phiplot} for $N = 7$, so we are comfortable with our assumption.

Next, we are expecting a non-perturbative function of the form $\Gamma/V \sim \exp\left[-F\left(N\right)\right]$ \cite{Witten:1980sp, Shifman:1998if}, so running simulations for $\Gamma/V\left(N\right)$ we plot the results in the form $F\left(N\right)$.
This plot is given in figure \ref{Fplot} where the points and error bars given are the mean and standard deviation for $25$ trials of our simulation at each $N$ between $15$ and $75$ using $312$ Chebyshev grid points; it is shown on a log-log scale to emphasize the power law behaviour of $F\left(N\right)$.

The particular fit parameters are given for completeness but should not be regarded as terribly important.
In fact, the most important result for the present analysis  is that the computations are performed in theoretically controllable way where every single step is theoretically justified  in semi-classical limit (\ref{debye}) governed by the parameter ${\cal N}\gg 1$. 
We are after all working in a toy model, and as such should expect a good qualitative picture but not take the numerical details too seriously.
It is however interesting that the final form for the decay rate is given as, putting the parameter ${\cal N}$ back in,
\be	\label{finalresult}
  \frac{\Gamma}{V} \sim \exp \left\lbrace -{\cal N}\left(aN^b\right) \right\rbrace
\ee
with both $a$ and $b$ positive.
Thus, the decay rate does drop off exponentially in $N$ and our other semiclassical parameter ${\cal N}$, and indeed faster than any perturbation term would describe as previously conjectured. 
It is a semiclassical calculation, but the behaviour is fundamentally non-perturbative, and it is only parametrically justified when ${\cal{N}}\gg 1$. 

A few comments are in order.
First, our numerical estimates (\ref{finalresult}) can be only trusted for finite $N>5$, but not for parametrically large $N\rightarrow\infty$ where the region of validity of the model shrinks to a point, see footnote \ref{N}.
Furthermore, if the external parameter $N$ were allowed to vary in a very large region it may lead (and, in fact, it does) to a systematic error in our numerical simulations.
This is because in our numerical simulations we assume that all our variables are order of unity, rather than having some functional dependence on $N$, which may not be the case when the external parameter $N$ is allowed to vary in wide region of parameter space. 
Finally, one should not expect that our formula (\ref{finalresult}) would reproduce the asymptotic behaviour \cite{Shifman:1998if} due to the differences in large $N$ scaling in ``deformed QCD'' model  and in  strongly coupled QCD, see footnote \ref{N}.
As mentioned previously, the main goal of our computations is to support  the  qualitative, rather than quantitative picture of metastable vacua and their decay, conjectured in \cite{Witten:1980sp,wittenflux} in a simplified model where calculations are parametrically justified at ${\cal{N}}\gg 1$ and finite $N$.
Nevertheless, there is a room to improve our numerical simulations in a much wider range of $N$ as a result of the recent analysis \cite{Li:2014} in which the asymptotic expression at $N\rightarrow\infty$ has been analytically computed.
These improvements are discussed in the next section.

\subsection{Improved Results}	\label{improved}

Recently, an analytical analysis of the asymptotic behaviour of this calculation, inspired by the above numerical results, has been carried out \cite{Li:2014} with the asymptotic expression for the decay rate per unit volume given by
\be	\label{asymptotic}
  \frac{\Gamma}{V} \sim \exp \left[-\mathcal{N}\frac{256 N^{7/2}}{9\sqrt{3}\pi\left(\pi-\theta\right)^2}\right].
\ee
The asymptotic expression (\ref{asymptotic}) gives us a hint about how to produce a better estimate for the decay rate for very large $N \gg 1$ in comparison with our naive numerical results presented in section \ref{results} above, wherein we assumed that parameter $N$ is not allowed to vary in an extended region of parameter space.

Indeed, the analysis in \cite{Li:2014} suggests the specific reason for the disparity between the asymptotic expression and the numerical results shown in figure \ref{Fplot} for large $N$.
Mainly, the asymptotic guess solution is given by
\be	\label{asymptoticguess}
  \sigma_n\left(x\right) = \left(\frac{4n}{N}-2\right)\arctan\left[\exp\left(-2\sqrt{\frac{3}{N}}x\right)\right],
\ee
which has a size $\sim\!\sqrt{N}$ changing with the parameter $N$. 
Our guess solution (\ref{guess}) does not  scale  with $N$ and so becomes an increasingly bad guess at asymptotically larger $N$, such that our numerical solver becomes increasingly likely to find some other local minimum of the action.
Furthermore, our integration domain was fixed for all $N$ as we did not even attempt to consider any large variations with $N$ in our analysis in section \ref{results}.
When we allow the external parameter $N$ to become ``large'', the true minimal action interpolating trajectory eventually will not ``fit'' in the finite size numerical grid which we fixed for all $N$.
Essentially, forcing boundary conditions on too small a domain also forces a higher action local minimum as $N$ increases.
This is precisely the mechanism by which a systematic error is introduced into the numerical simulations as a result of large variation in the external parameter $N$, as suggested in the previous section after (\ref{finalresult}).

Fortunately, the analytical expression (\ref{asymptoticguess}) which is valid for asymptotically large $N$ suggests a simple fix to improve our numerical solution at higher $N$ by explicitly taking into account the variation of the trajectory size with this parameter.
Technically, we can allow the integration domain to scale $\sim\!\sqrt{N}$, and start with the asymptotic guess (\ref{asymptoticguess}) in which the large parameter $N$ explicitly enters, as the initial guess for the ``improved'' numerical algorithm.
Again, we added some Gaussian noise to get an ensemble of $25$ initial guesses for each $N$ and relaxed them as described in Section \ref{numerical} to arrive at a minimum of the action.
These improved results are shown in figure \ref{modFplot} plotted along with the asymptotic expression for the decay rate and the first few points from the original simulations in Section \ref{results}.
They reproduce the previous results for finite $N \leq 15$ and approach the asymptotic result (\ref{asymptotic}) given in \cite{Li:2014} from below for large $N$.
Numerically, the asymptotic expression (\ref{asymptotic}), which is formally valid at $N\rightarrow\infty$, describes our improved simulation data sufficiently well (with accuracy better than $10\%$) only at large $N \geq 35$.

\begin{figure*}
  \begin{center} 
    \includegraphics[width = \textwidth]{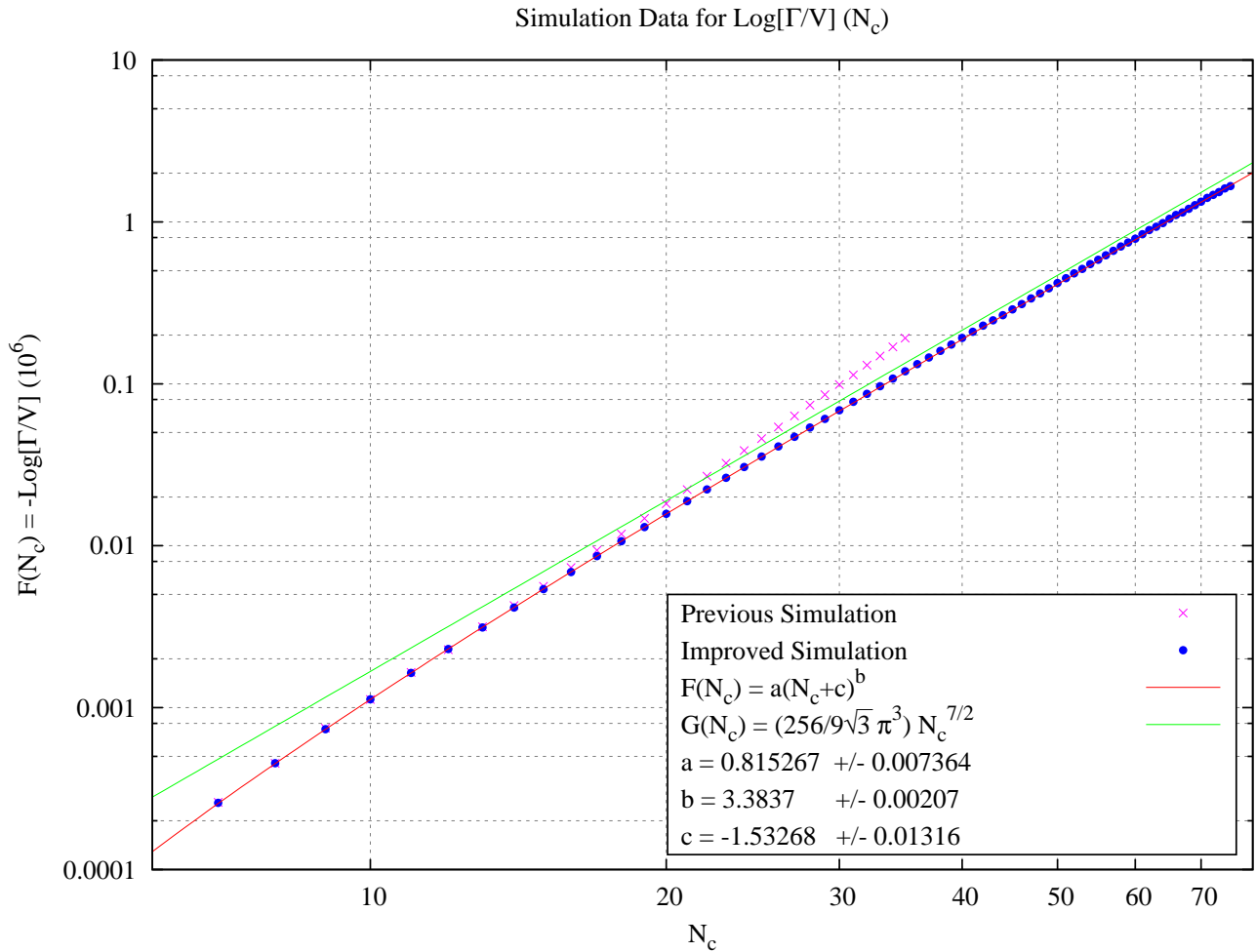}
      \caption{		\label{modFplot}
	A plot of the improved simulation data for the decay exponent $F\left(N\right)$ plotted for $N$ in the range $7$ to $75$.
      }
  \end{center}
\end{figure*}

\section{Conclusion}		\label{conclusion}
Our conclusion can be separated on two different parts: \\1). solid theoretical results within  the ``deformed QCD" model, and \\ 2). some  speculations related to strongly coupled QCD realized in nature.

We start with the first part of the conclusion in which our basic result is as follows.
We have demonstrated that the ``deformed QCD'' model shows (once again) that some qualitative features expected to occur in strongly coupled regime in large $N$ limit as argued in \cite{Witten:1980sp} do emerge in the simplified version of the theory as well.
We demonstrated the existence of metastable vacuum states with energy density higher than the ground state by $\epsilon \sim 1/N$, and have shown that the life time of the metastable states is exponentially suppressed in this model with respect to  semi-classicality parameter ${\cal N}$.
The suppression increases even further with increasing number of colours $N$ for a fixed ${\cal N}$, and it is given by (\ref{finalresult}). 

In this simplified system one can explicitly see these metastable states, how they are classified, and the microscopic dynamics which govern the corresponding physics.
We shall not repeat this analysis, instead referring to section \ref{classification}.
However, the important remark we would like to make here which is relevant for the speculative portion of our conclusion is that the  $\cal{P}$ and $\cal{CP}$ invariance is generally violated in these metastable vacuum states as the expectation value for the topological density (\ref{top}) explicitly shows. 
We believe that this feature of spontaneous breaking of the $\cal{P}$ and $\cal{CP}$ invariance in metastable states is quite a generic feature which is shared by strongly coupled pure gauge theories (for sufficiently large $N$).
Precisely this feature of the metastable states plays a crucial role in our speculative portion of the conclusion.

Therefore, we now speculate that precisely this spontaneous symmetry breaking effect is responsible for the asymmetries in event by event studies observed at the RHIC (Relativistic Heavy Ion Collider) and the LHC (Large Hadron Collider). 
To be more specific, the violation of local $\cal{P}$ and $\cal{CP}$ symmetries has been the subject of intense studies for the last couple of years as a result of very interesting ongoing experiments at RHIC  \cite{Abelev:2009tx,Wang:2012qs} and, more recently, at the LHC   \cite{Selyuzhenkov:2012py,Abelev:2012pa,Voloshin:2012fv,Selyuzhenkov:2012mf}, see \cite{Kharzeev:2013ffa} for a recent review and introduction to the subject with a large number of references to original papers.

The main idea for explaining the observed asymmetries is to assume \cite{Kharzeev:2007tn,Kharzeev:2013ffa} that an effective $\theta (\vec{x}, t)_{ind} \neq 0$ is induced in the process of cooling of the system representing the high temperature quark-gluon plasma.
In other words, the system in the process of cooling may spontaneously choose one or another state which is not the absolute minimum of the system corresponding to the $\theta=0$, but rather, some excited state, similar to the old idea when the disoriented chiral condensate can be formed as a result of heavy ion collisions.
The key assumption is that this induced $\theta (\vec{x}, t)_{ind} \neq 0$ is coherent on a large scale of order of nuclei $\sim 10$~fm, rather than on much smaller scale of order of 1~fm.
If a state with $\la\theta (\vec{x}, t)_{ind}\ra\neq 0$ is indeed induced, it implies a violation of the local ${\cal{P}}$ and ${\cal{CP}}$ symmetries on the same scales where $\theta (\vec{x}, t)_{ind} \neq 0$  is  correlated.
It may then generate a number of ${\cal{P}}$ and ${\cal{CP}}$ violating effects, such as Charge/Chiral Separation (CSE) and Chiral Magnetic (CME) Effects, see \cite{Kharzeev:2013ffa} for a recent review. 

One of the critical questions for the applications of the CME to heavy ion collisions is a correlation length of the induced $\la\theta (\vec{x}, t)_{ind}\ra \neq 0$. 
Why are these ${\cal{P}}$ odd domains large?

We would like to speculate that the crucial element in the understanding of this key question might be related to the metastable states which are the subject of the present work.
To be more specific, we suggest that the system being originally formed at high temperature might be locked in one of these metastable states during the cooling stage\footnote{\label{asymmetry}The ${\cal{P}}$ and ${\cal{CP}}$ symmetries, of course, are good symmetries in QCD. The probability to produce $m=+ 1$ state  from eq. (\ref{sufficient}) is identically the same as produce $m=- 1$ state. Therefore, there will be no any  ${\cal{P}}$ and ${\cal{CP}}$  violating effects  if one averages  over large number of events. However, one should expect some asymmetries if one analyzes  the system on event by event basis, which is precisely the procedure used at RHIC and the LHC, see  ref. \cite{Kharzeev:2013ffa} for a recent review.}.
If this happens one should obviously expect a number of ${\cal{P}}$ and ${\cal{CP}}$ effects to occur coherently in the entire system characterized by a large scale of order the size of nuclei $\mathbb L\gg \Lqcd^{-1}$. 
We therefore identify $\theta (\vec{x}, t)_{ind} \neq 0$ from \cite{Kharzeev:2007tn} with the effective theta parameter $2\pi/N$ which enters (\ref{top}) and which manifests a spontaneous violation of the  ${\cal{P}}$  and ${\cal{CP}}$ symmetries in the system. 

The presence of such long range order (which itself is a consequence of a spontaneous selecting of a metastable vacuum state in the entire system during the cooling process) may explain why CME is operational in this system and how the asymmetry can be coherently accumulated from entire system.
This identification would justify the effective Lagrangian approach advocated in \cite{Kharzeev:2007tn,Zhitnitsky:2012im} wherein $\theta (\vec{x}, t)_{ind} $ is treated as slow background field with correlation length much larger than any conventional QCD fluctuations, $\mathbb L\gg \Lqcd^{-1}$.
It is important to emphasize that the ${\cal{P}}$ and ${\cal{CP}}$ symmetries are good symmetries of the fundamental QCD.
As mentioned in footnote \ref{asymmetry} the asymmetries can only be observed in heavy ion collisions in event by event analysis when the system might be locked for sufficiently long period of time $\tau\sim \mathbb L/c\gg \Lqcd^{-1}$ in a metastable state in one collision with one specific sign for the topological density (\ref{top}).
Because the metastable states with opposite signs for the topological density operator (\ref{top}) have the same energy, which state is chosen for a particular event is random and evenly distributed.
Thus, it is clear that if one averages over large number of events, the asymmetry will be washed out as the probability to form these metastable states is identical and the lifetime for the two is the same as we mentioned in section \ref{classification}.
However, in the event by event studies the asymmetry will be evident in the system.
Apparently, this is precisely what has been observed, see the recent review paper \cite{Kharzeev:2013ffa} for the details. 

\section*{Acknowledgements}

E. Thomas would like to thank Michael Forbes and Darren Smyth for discussions related to the solving of nonlinear boundary value problems.

This research was supported in part by the Natural Sciences and Engineering Research Council of Canada.


\end{document}